\documentclass[notitlepage,twoside,11pt,letter]{article}

\usepackage{times}
\usepackage{amsmath}
\usepackage{amsfonts}
\usepackage{amssymb}
\usepackage{graphicx}
\usepackage{mathrsfs}
\usepackage{tabularx}
\usepackage{caption}
\usepackage{bbold}
\usepackage{color}
\usepackage{fancybox}
\usepackage{verbatim}
\usepackage{hyperref}
\usepackage{tikz}
\usetikzlibrary{arrows,automata}
\usetikzlibrary{trees}

\def\b1{\mathbb{1}}

\def\eE{\mathsf{E}}

\def\tT{\mathtt{T}}

\def\e1{\mathsf{1}}

\def\of0{(0)}

\def\bf0{\mathbf{0}}
\def\cp1{\mathbb{CP}^1}

\begin{document}

\title{\textbf{Options on infectious diseases}}
\author{\textbf{Andrew Lesniewski and Nicholas Lesniewski}\\
Department of Mathematics\\
Baruch College\\
55 Lexington Avenue\\
New York, NY 10010
\date{Draft of \today\\
First draft: March 17, 2020}}
\maketitle

\begin{abstract}
We present a parsimonious stochastic model for valuation of options on the fraction of infected individuals during an epidemic. The underlying stochastic dynamical system is a stochastic differential version of the SIR model of mathematical epidemiology.
\end{abstract}

\section{Introduction}\label{introSec}

In the current environment of economic uncertainty related to the spread of COVID-19, it could be of interest to develop financial instruments that allow market participants to express their views on the economic impact of an epidemic. The simplest such instruments are \textit{infection options}, which are options on the fraction of infected individuals in a population affected by an epidemic\footnote{According to \url{https://www.worldometers.info/}, as of March 17, the fractions of populations currently infected by COVID-19 is 0.0018\% for the USA, 0.0313\% for Switzerland, and 0.0431\% for Italy.}. Infection options would allow dynamic hedging of risk exposure to pandemic linked catastophe bonds. Investors purchasing catastrophe bonds provide insurance to issuers and lose invested princpal if certain conditions are met\footnote{In June 2017 the World Bank issued pandemic linked catastrophe bonds and derivatives worth \$425 million.}.


In this note we propose a mathematical model for pricing and risk management of such options. One of the challenges is to formulate a model which, at the very least, exhibits typical stylized facts about the spread of highly infectious diseases such as COVID-19, while remaining numerically tractable, intuitive, and interpretable. The basic category of models of mathematical epidemiology \cite{BCF19} are compartmental models, which divide populations into homogeneous compartments. The model we selected is a stochastic differential equation extension of a member of that family of models, namely the Susceptible-Infected-Recovered (SIR) model.

The SIR model of infectious diseases is among the simplest models, and was originally published in 1927 by Kermack and McKendrick \cite{KM27}. The classic SIR model is formulated in terms of three state variables:
\begin{itemize}
\item[(i)]{$0\leq x\leq 1$, the fraction of individuals who are \textit{susceptible} to the disease,}
\item[(ii)]{$0\leq y\leq 1$, the fraction of individuals who are \textit{infected} with the disease,}
\item[(iii)]{$0\leq z\leq 1$, the fraction of individuals who have \textit{recovered} and are immune to the disease.}
\end{itemize}
The model dynamics is given as the following dynamical system:
\begin{equation}
\begin{split}
\frac{\dot x(t)}{x(t)}&=-\beta y(t),\\
\frac{\dot y(t)}{y(t)}&=\beta x(t)-\gamma,\\
\frac{\dot z(t)}{y(t)}&=\gamma,
\end{split}
\end{equation}
with the initial condition:
\begin{equation}\label{initcond}
\begin{split}
x(0)&=x_0,\\
y(0)&=y_0,\\
z(0)&=1-x_0-y_0.
\end{split}
\end{equation}
The constant parameters $\beta>0$ and $\gamma>0$ are called the infection and recovery rates, respectively.

Notice that this dynamics obeys the conservation law
\begin{equation}\label{sumofvarbs}
x(t)+y(t)+z(t)=1,
\end{equation}
consistent with the assumption that the variables $x,y$, and $z$ represent  population fractions. This means that the variable $z$ is, in a way, redundant, as its current value does not affect the dynamics of $x$ and $y$, and it can be computed in a straightforward manner from \eqref{sumofvarbs}.

\section{Stochastic SIR models}

Stochastic extensions of the SIR model go back to Bartlett \cite{B56}, who formulated a stochastic jump process describing the evolution of an epidemic. Various other stochastic extensions have been proposed since, see e.g. \cite{B10}. Here, we will focus on two stochastic extensions of the SIR model, which utilize stochastic differential equations, see e.g. \cite{GGHMP11} and references therein. These models are particularly adept at option modeling.

The first of the models, a one factor stochastic SIR model is obtained in the following manner. Let $W_t$ denote the standard Brownian motion, and let $\dot W_t$ denote the (generalized) white noise process. We assume that the infection rate $\beta$, rather than being constant, is subject to random shocks, namely $\beta_t=\beta+\sigma\dot W_t$, while the recovery rate $\gamma$ remains constant. Here $\sigma>0$ is a constant volatility parameter. This leads to the following stochastic dynamics:
\begin{equation}\label{onefactmod}
\begin{split}
\frac{d X_t}{X_t}&=-\beta Y_t dt-\sigma Y_t dW_t,\\
\frac{d Y_t}{Y_t}&=(\beta X_t-\gamma)dt+\sigma X_t dW_t,\\
\frac{d Z_t}{Y_t}&=\gamma dt,
\end{split}
\end{equation}
with the initial condition analogous to \eqref{initcond}.

Notice that the conservation law
\begin{equation}\label{stochconslaw}
X_t+Y_t+Z_t=1
\end{equation}
continues to hold in the stochastic model, and so we can focus attention on the dynamics described by $X$ and $Y$ only.

Consider now a time horizon $T>0$, and $\xi,\eta>0$. The backward Kolmogorov equation for the Green's function $G=G_{T,\xi,\eta}(t,x,y)$ corresponding to the stochastic system \eqref{onefactmod} reads:
\begin{equation}\label{bke1}
\dot{G}-\beta xy\nabla_x G+(\beta x-\gamma)y\nabla_y G+\frac12\sigma^2 x^2 y^2\big(\nabla_x^2 G-2\nabla^2_{xy}G+\nabla_y^2 G\big)=0.
\end{equation}
We assume that $G_{T,\xi,\eta}(t,x,y)$ satisfies the following terminal condition:
\begin{equation}\label{termcond}
G_{T,\xi,\eta}(T,x,y)=\delta(x-\xi)\delta(y-\eta).
\end{equation}

Similarly, a two factor stochastic SIR model can be formulated in the following manner. In addition to allowing the infection rate being stochastic, we assume that the recovery rate is subject to random shocks delivered by a second Brownian motion $B_t$, $\gamma_t=\gamma+\zeta\dot B_t$. Again, $\zeta>0$ is a volatility parameter. The two Brownian motions are not required to be independent, and we let $\rho$ denote the correlation coefficient between them,
\begin{equation}\label{corr}
dW_t dB_t=\rho dt.
\end{equation}
This leads to the following stochastic dynamics:
\begin{equation}\label{twofactmod}
\begin{split}
\frac{d X_t}{X_t}&=-\beta Y_t dt-\sigma Y_t dW_t,\\
\frac{d Y_t}{Y_t}&=(\beta X_t-\gamma)dt+\sigma X_t dW_t-\zeta dB_t,\\
\frac{d Z_t}{Y_t}&=\gamma dt+\zeta dB_t,
\end{split}
\end{equation}
with the initial condition analogous to \eqref{initcond}. The two factor stochastic model continues to obey the conservation law \eqref{stochconslaw}. Notice that the two factor model has two additional parameters compare to the one factor model, namely the volatility of the recovery rate $\zeta$ and the correlation coefficient $\rho$ in \eqref{corr}.

The backward Kolmogorov equation for the stochastic system \eqref{twofactmod} reads as follows. Set
\begin{equation}\label{eta}
\eta(x)=\sqrt{\sigma^2 x^2-2\rho\sigma\zeta x+\zeta^2}.
\end{equation}
Then
\begin{equation}\label{bke2}
\begin{split}
\dot{G}&-\beta xy\nabla_x G+(\beta x-\gamma)y\nabla_y G\\
&+\frac12y^2\Big(\sigma^2 x^2\nabla_x^2 G-2\sigma x(\sigma x-\rho\zeta)\nabla^2_{xy}G+\eta(x)^2\nabla^2_y G\Big)=0,
\end{split}
\end{equation}
where the Green's function $G=G_{T,\xi,\eta}(t,x,y)$ satisfies terminal condition \eqref{termcond}.

\section{Infection options}

The payoff of the call option on the infection fraction struck at $K$ is given by $(Y_T-K)^+$ multiplied by the notional amount on the contract. Here, as usual, $x^+=\max(x,0)$. Likewise, the payoff of the put option on the infection fraction struck at $K$ is $(K-Y_T)^+$ multiplied by the notional amount on the contract

Infection options are written on an underlying which is not a financial asset. For this reason, they are much closer in spirit to weather options (or real options), than to traditional equity or FX options. Consequently, there is no natural risk neutral approach to their valuation.

For this reason, we adopt an approach that is based on the process \eqref{onefactmod} (or, alternatively, the more complicated process \eqref{twofactmod}), which is written under the physical measure. Today's value of an infection call option is thus given by
\begin{equation}
\mathrm{Call}(T,K)=e^{-rT}\eE((Y_T-K)^+)\times\text{ notional amount},
\end{equation}
where $r$ is the discount rate, and $\eE$ denotes expected value with respect to the probability measure generated by the process \eqref{onefactmod} (or \eqref{twofactmod}). In terms of the Green's function $G$, this can be expressed as a two dimensional integral over the terminal probability distribution:
\begin{equation}
\eE((Y_T-K)^+)=\iint_{[0,1]^2} G_{T,\xi,\eta}(0,X_0,Y_0)(\eta-K)^+d\xi\,d\eta.
\end{equation}
Note that while the current value of the option depends on the current reading of the susceptible fraction, its value at expiration is not set. Analogous formulas hold for put options. 

Closed form solutions to equations \eqref{bke1} and \eqref{bke2} are not available and the evaluation of the expected values requires numerical computations. Two methods are of practical importance.

Substituting $t\to T-t$ we convert the (two-dimensional) backward Kolmogorov equation to a diffusion equation, which can be solved through a number of standard methods, such as the ADI method, see e.g. \cite{MM05}.

Alternatively, one can calculate the expected values using Monte Carlo simulations \cite{G03}. Specifically, a convenient discretization to \eqref{onefactmod} is formulated as follows. In order to guarantee the positivity of $X_t$ and $Y_t$ throughout the simulation, we set $X_t=\exp(-l_t)$ and $Y_t=\exp(-m_t)$. Denoting by $\delta$ the magnitude of the time step, we are lead to the following Euler scheme:
\begin{equation}
\begin{split}
l_{n+1}&=l_n+(\beta\delta+\sigma \sqrt{\delta}\,z_n)e^{-m_n}+\frac12\,\sigma^2 \delta e^{-2m_n},\\
m_{n+1}&=m_n+\gamma\delta-(\beta\delta+\sigma \sqrt{\delta}\,z_n)e^{-l_n}+\frac12\,\sigma^2 \delta e^{-2l_n},\\
\end{split}
\end{equation}
for $n=0,1,\ldots$. Here, $z_n$ are independent variates drawn from the standard normal distribution. At each iteration step one has to floor $l_{n+1}$ and $m_{n+1}$ at $0$, $l_{n+1}=\max(l_{n+1},0)$, $m_{n+1}=\max(m_{n+1},0)$, so that $X_t, Y_t\leq 1$.

Similarly, the two factor system \eqref{twofactmod} can be discretized as follows:
\begin{equation}
\begin{split}
l_{n+1}&=l_n+(\beta\delta+\sigma \sqrt{\delta}\,z_n)e^{-m_n}+\frac12\,\sigma^2 \delta e^{-2m_n},\\
m_{n+1}&=m_n+\gamma\delta-(\beta\delta+\sigma \sqrt{\delta}\,z_n)e^{-l_n}+\zeta \sqrt{\delta}\,b_n+\frac12\,\eta(e^{-l_n})^2\delta,\\
\end{split}
\end{equation}
with $\eta(x)$ defined in \eqref{eta}, for $n=0,1,\ldots$. The two dimensional random normal vectors $(z_n, b_n)^\tT$ are independently drawn from the Gaussian distribution with covariance
\begin{equation}
C=
\begin{pmatrix}
1&\rho\\
\rho&1
\end{pmatrix}.
\end{equation}

\end{document}